\documentclass{article}

\PassOptionsToPackage{numbers, numbers}{natbib}

\usepackage[final]{neurips_2024}

\usepackage[utf8]{inputenc} % allow utf-8 input
\usepackage[T1]{fontenc}    % use 8-bit T1 fonts
\usepackage{hyperref}       % hyperlinks
\usepackage{url}            % simple URL typesetting
\usepackage{booktabs}       % professional-quality tables
\usepackage{amsfonts}       % blackboard math symbols
\usepackage{nicefrac}       % compact symbols for 1/2, etc.
\usepackage{microtype}      % microtypography
\usepackage{xcolor}         % colors
\usepackage{amsmath}
\usepackage{graphicx}
\usepackage{algorithm}
\usepackage{algorithmic}

\title{Learning Pore-scale Multi-phase Flow from Experimental Data with Graph Neural Network}

\author{%
  Yuxuan Gu \\
  Department of Electrical and Electronic Engineering \\
  Imperial College London\\
  \texttt{yuxuan.gu21@imperial.ac.uk} \\
  \And
  Catherine Spurin \\
  Department of Energy Science and Engineering \\
  Stanford University \\
  \texttt{cspurin@stanford.edu} \\
  \And
  Gege Wen \\
  Department of Earth Science \& Engineering, I-X \\
  Imperial College London\\
  \texttt{g.wen@imperial.ac.uk} \\
}

\begin{document}

\maketitle

\begin{abstract}
Understanding the process of multiphase fluid flow through porous media is crucial for many climate change mitigation technologies, including CO$_2$ geological storage, hydrogen storage, and fuel cells. 
However, current numerical models are often incapable of accurately capturing the complex pore-scale physics observed in experiments. 
In this study, we address this challenge using a graph neural network-based approach and directly learn pore-scale fluid flow using \textit{micro-CT experimental data}. We propose a Long-Short-Edge MeshGraphNet (LSE-MGN) that predicts the state of each node in the pore space at each time step. During inference, given an initial state, the model can autoregressively predict the evolution of the multiphase flow process over time. This approach successfully captures the physics from the high-resolution experimental data while maintaining computational efficiency, providing a promising direction for accurate and efficient pore-scale modeling of complex multiphase fluid flow dynamics.

\end{abstract}

\section{Introduction}
Our society is facing unprecedented challenges in climate change, demanding swift action to accelerate the energy transition toward net-zero~\cite{IPCC2018}. 
Understanding the process of multiphase flow through porous media is an important task as it is involved in many mitigation technologies, including CO$_2$ geological storage (\cite{wilberforce2021progress, ma2022carbon, shu2023role}), hydrogen storage~\cite{muhammed2022review}, and fuel cells~\cite{andersson2016review}. To study these multiphase flow processes, scientists can utilize micro-CT scanners with synchrotron sources to image the rock pores in a nanometer-scale spatial resolution while recreating the in situ condition of gas flowing through the porous medium~\cite{blunt2013pore}. State-of-the-art research facilities can now generate 3D imaging data with billions of voxels with temporal resolution on the order of second~\cite{spurin2020real}, creating a unique opportunity to advance our understanding of fluid flow physics. 

% Predicting fluid distribution in multiphase flow is essential for the successful engineering of these tasks. To study these processes, we can run experiments... Routinely obtain billions of voxel data... Provides a unique opportunity to study fluid flow xxx and advance fundamental learning of the physics...

% The micro-CT experiments 
However, despite the advancement in experimental capability to obtain high-resolution experimental data, the modeling of pore-scale multiphase flow in porous media remains very challenging. Current modeling approaches generally fall into three major classes: lattice-based models, continuum models, and pore-network models~\cite{zhao2019comprehensive}. The former two approaches incorporate the Navier-Stokes equation to solve flow equations but are often highly computationally expensive~\cite{gackiewicz2021intercomparison}. Pore network models simplify porous media into interconnected pores and throats~\cite{xiong2016review}, offering computational efficiency but often lacking accuracy due to the complexity of the physics involved.

Machine learning-based approaches are emerging in recent pore-scale modeling studies~\cite{santos2020poreflow, zhou2022neural}. However, most existing studies aim to learn fluid flow physics from simulation data, which suffers from significant drawbacks. Due to the aforementioned computational challenges in pore-scale modeling, training datasets are often incapable of reflecting the accurate behavior of real-world multiphase flow. They are often simulated with highly simplified pore geometry (e.g., packed spheres~\cite{zhou2022neural}) or simplified physics (e.g., negligible viscous effects and assumptions of steady-state flow). In addition, many previous approaches were designed based on convolutional neural networks (CNNs) that are optimized for grid-like structures. As a result, they struggled with the irregular geometries of complex pore structures (see example in Appendix A). 

% For example, Santos \textit{et al.} proposed PoreFlow-Net, a convolutional neural network-based model to predict single-phase flow in porous rocks \cite{santos2020poreflow}. 
% While PointNet++ \cite{qi2017pointnet++} highlights the effectiveness of 3D convolution on point clouds, 
% For instance, CNNs 

\begin{figure}[!b]
  \centering
  \includegraphics[width=\textwidth]{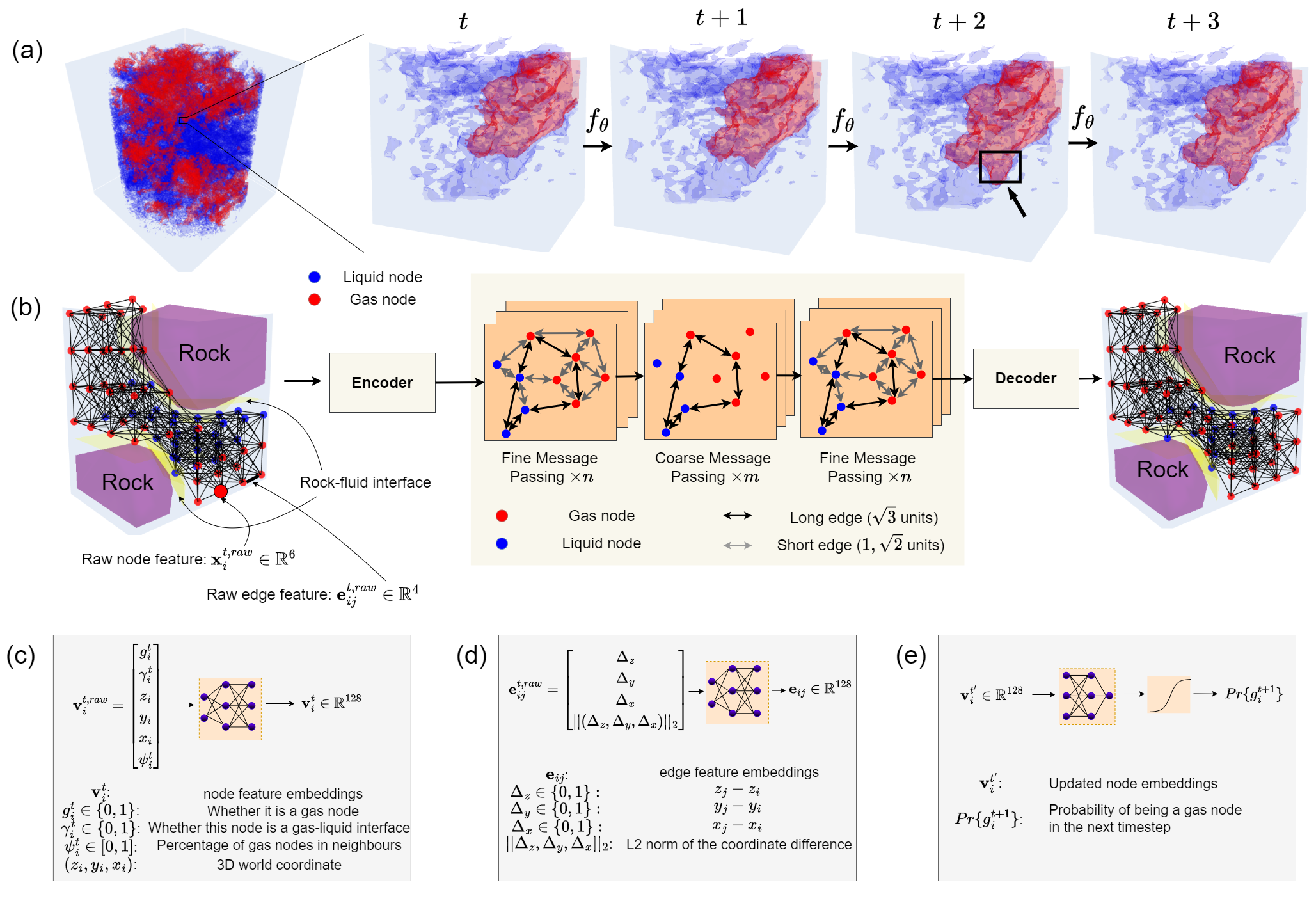}
  \caption{(\textbf{a}) A section in the rock is used to train a GNN model $f_{\theta}$. It learns to predict the next state based on its current state. (\textbf{b}) Five steps in the prediction pipeline: step 1. encoder that embeds node and edge features into latent space; step 2. fine message passing with all edges; step 3. coarse message passing only with long edges; step 4. fine message passing with all edges; step 5. a decoder that outputs the final node states. (\textbf{c}) Node encoder. (\textbf{d}) Edge encoder. (\textbf{e}) Decoder.}
  \label{gnn_model}
\end{figure}

To address the challenges, here we propose the Long-Short-Edge MeshGraphNet (LSE-MGN), a graph neural network (GNN)--based architecture for learning pore-scale fluid flow directly from experimental data (Figure~\ref{gnn_model}). Inspired by the \textsc{MeshGraphNets}~\cite{pfaff2020learning} architecture proposed by Pfaff \textit{et al.}, we used graph structure to represent irregular pore structures, where nodes represent fluid volume and edges represent fluid flow paths. Unlike CNNs, the graph-based representation can handle complex geometries by passing messages only between connected nodes, ignoring nodes separated by solid rock grains. Learning the fluid flow dynamics using a GNN-based architecture leverages inductive biases to focus on local interactions between fluid and pore structures. This allows for zero-shot generalisation to other boundary conditions~\cite{sanchez2020learning} or bigger input fields through section-based training~\cite{wu2022learning}. Learning directly from experimental data allows us to replicate realistic flow phenomena, including those that remain beyond the reach of simulations due to incomplete understanding. Our results demonstrate that LSE-MGN can effectively capture complex fluid behaviors and generalise well across varying boundary conditions. Our approach provides a new paradigm to leverage the extensive micro-CT experimental data to study pore-scale physics.

%  developed \textsc{MeshGraphNets} \cite{pfaff2020learning}, which use mesh-based data, allowing for flexible resolution allocation in different regions, which is a challenge for particle- and grid-based simulations.

% The nodes in GNS represent particles in the simulation space and the edge represents the interaction with neighbourhood particles.

% In MGN, vertices of the mesh are represented as nodes, and edges are derived from cells shaped as \texttt{(timestep, cell\_num, 3)}. Each cell includes the indices of three vertices forming it. 

% Previous studies have used simulation data to train machine learning models, but these often rely on simplifications , which do not fully capture real-world conditions. Tracking individual particle velocities in experiments is also challenging. Machine learning methods have not yet been applied to predict multiphase fluid flow using real experimental data in CCS, highlighting a significant research gap. 

\section{Data Description}\label{data_description}
The micro-CT multiphase flow experiment dataset used in here was collected at the Swiss Light Source, Paul Scherrer Institute. During the experiment, the rock sample was exposed to filtered polychromatic X-ray radiation originating from a 2.9 T bending magnet source, generating a spatial imaging resolution of 2.75 $\mu m$ and a temporal resolution of 2 $s$. The experiments were conducted in a cylindrical carbonate sample, 5~mm diameter and 20~mm length. The sample was saturated with brine (deionised water doped with 15wt.\% KI to improve the X-ray contrast). Then nitrogen and brine were injected simultaneously. X-ray images were acquired to capture the transition from transient effects to steady-state flow. Each image took 2~s to acquire, and had a resolution of 2.75~$\mu m$. For a complete experimental description see \cite{spurin2020real}. Steady-state was determined by a stable pressure drop measured across the core. The greyscale X-ray images were segmented to extract the location of the fluids present (gas and brine). This was done using the method described in \cite{spurin2023python}. The dataset was published in Spurin \textit{et al.} \cite{spurin2020real}. 

In this work, we focus on a $50 \times 80 \times 50$ cube over 300-time steps characterised by highly intermittent fluid flow, where the flow pathways through the pore space are constantly rearranging (Figure~\ref{gnn_model}~(a)). Fluid intermittency is a phenomenon that current numerical modeling techniques cannot accurately capture.

% Using \textsc{MeshGraphNets} within an \textit{Eulerian}\footnote{\textit{Eulerian} systems model continuous fields over a fixed mesh, sampling at mesh nodes \cite{pfaff2020learning}. } framework, we model fluid dynamics in pore spaces to predict fluid distribution at the next time step. Autoregressive inference generates a sequence of future states, as depicted in Figure \ref{gnn_model}(a). In the raw data, each 3D scan has a shape of $1550 \times 1563 \times 1536$, resulting in 424,535,514 nodes in the pore space. To overcome the memory constrain for the large data size, we used section-based training approach proposed in~\cite{ \cite{wu2022learning}}, where we divide the sections into $50 \times 80 \times 50$ cubes for training.

% Authors in \cite{pfaff2021learning} mentioned two systems: \textit{Eulerian} and \textit{Lagrangian} systems. \textit{Eulerian} systems model continuous fields over a fixed mesh, sampling at mesh nodes. \textit{Lagrangian} systems represent a moving mesh with additional coordinates for its dynamic state in 3D space. 

\section{Methodology}\label{Methodology}

\paragraph{Graph construction.} We constructed the input and output graphs using the 3D images segmented into either rock or pore spaces. As shown in Figure~\ref{gnn_model} (b), the rock spaces were excluded from the graph, and the pore spaces were represented as interconnected nodes. At time step $t \in \{0,...,T\}$, each point $i$ in the pore space is represented as a node with a feature vector $\textbf{v}_i^{t}$ defined as

\begin{equation}
    \textbf{v}_i^{t} = [g_i^{t}, \gamma_i^{t}, z_i, y_i, x_i, \psi_i^{t}]^T,
\label{eq: node_attr}
\end{equation}

where $g_i^{t}\in \{0, 1\}$ indicate the state of a node (i.e., gas or liquid), and $\gamma_i^{t} \in \{0, 1\}$ indicates whether the node is at a gas-liquid interface. The spatial coordinates $\{z_i, y_i, x_i\}$ specify a node's position in the 3D pore space. $\psi_i^{t}$ represents the percentage of gas nodes among the one-hop neighbors of node $i$. We construct bidirectional edges by connecting nodes within a fixed radius $R$ of the current node. In this work, $R$ is set to $\sqrt{3}$ to ensure that message passing respects the pore structure. With this setup, a node can connect to a maximum of 27 surrounding nodes, including itself, as shown in Figure \ref{connectivity}~(b) in Appendix~\ref{pore_structure}. There are three distinct edge lengths, excluding the self-connection: 1, $\sqrt{2}$, and $\sqrt{3}$ units, respectively. Inspired by \cite{pfaff2020learning}, we defined the edge attribute $\textbf{e}_{ij}$ as

\begin{equation}
    \textbf{e}_{ij} = \begin{bmatrix}(z_i - z_j), (y_i - y_j), (x_i - x_j), ||((z_i - z_j), (y_i - y_j), (x_i - x_j))||_2\end{bmatrix}^T,
\label{eq: edge_attr}
\end{equation}

which include the coordinates difference between two nodes and the \textit{l}2 norm of the difference.

% Footnotes should be used sparingly.  If you do require a footnote, indicate
% footnotes with a number\footnote{Sample of the first footnote.} in the
% text. Place the footnotes at the bottom of the page on which they appear.
% Precede the footnote with a horizontal rule of 2~inches (12~picas).

% Note that footnotes are properly typeset \emph{after} punctuation
% marks.\footnote{As in this example.}

\paragraph{Model architecture.} We propose the Long-Short-Edge MeshGraphNet (LSE-MGN) architecture that predicts the gas/liquid distribution in pore spaces given the node and edge features at the previous time step. At test time, we autoregressively generate a sequence of future states given the previous state, as shown in Figure \ref{gnn_model}~(a). The model architecture is illustrated in Figure \ref{gnn_model}~(b), where we employ the "encoder-processor-decoder" structure as inspired by Graph
Network-based Simulators (GNS) \cite{sanchez2020learning} and MultiScale-MeshGraphNets \cite{fortunato2022multiscale}. 

In the encoder,  the node features are embedded in the latent vector of dimension 128 via a Multi-Layer Perceptron (MLP). Similarly, an edge encoder embeds the raw edge feature of dimension 4 to a latent vector of dimension 128, as shown in Figure \ref{gnn_model}~(d). The decoder predicts the fluid state at each node with a binary node-level classification, as illustrated in Figure \ref{gnn_model}~(e). The node embeddings are transformed and summarised into a scalar value, then passed through a Sigmoid function. The final output is the probability of the node being classified as a gas node.

In the multi-scale processor, we divided edges into two categories: long edge ($\sqrt{3}$) and short edge ($0, 1, \sqrt{2}$). Nodes at both ends of the long edges are defined as pivotal nodes. Initially, node and edge features are propagated along all edges to gather information from nearby nodes. Subsequently, information is transferred only between pivotal nodes along the long edges, allowing pivotal nodes to access information from distant nodes, thereby expanding their receptive fields. Finally, messages from the pivotal nodes are `shared' with nearby nodes by propagating messages along all edges again. This is achieved using $n$ fine processors that aggregate features along all edges, then $m$ coarse processors that propagate messages only along the long edges, and then another $n$ fine processors along all edges. The fine message-passing processor with residual connections is defined as, 

% \begin{equation}
%     \textbf{e}_{ij}' \leftarrow \textbf{e}_{ij} + f^W \left(  \textbf{e}_{ij}, \textbf{v}_i, \textbf{v}_j \right) , \quad \textbf{v}_i' \leftarrow \textbf{v}_i + f^V ( \textbf{v}_i, \sum_{j \in \mathcal{N}(i) } \textbf{e}_{ij}' ) , 
% \label{fine mp}
% \end{equation}

% \begin{equation}
%     \textbf{e}_{ij}' \leftarrow \textbf{e}_{ij} + f^W (  \textbf{e}_{ij}, \textbf{v}_i, \textbf{v}_j ) \cdot \mathbb{I}(|\textbf{e}_{ij}| = \sqrt{3}), \quad \textbf{v}_i' \leftarrow \textbf{v}_i + f^V ( \textbf{v}_i, \sum_{j \in \mathcal{N}(i) } \textbf{e}_{ij}' )\cdot \mathbb{I}(|\textbf{e}_{ij}'| = \sqrt{3}), 
% \label{coarse mp}
% \end{equation}

\begin{equation}
 \textbf{e}_{ij}' \leftarrow \textbf{e}_{ij} + f^W \left(  \textbf{e}_{ij}, \textbf{v}_{i}, \textbf{v}_{j} \right), \quad \textbf{v}_i' \leftarrow \textbf{v}_i + f^V( \textbf{v}_i, \sum_{j \in \mathcal{N}(i)} \textbf{e}_{ij}'),
\label{fine mp}
\end{equation}

and the coarse processor defined as,

\begin{equation}
\textbf{e}_{ij}' \leftarrow \textbf{e}_{ij} + f^W (  \textbf{e}_{ij}, \textbf{v}_i, \textbf{v}_j ) \cdot \mathbb{I}(|\textbf{e}_{ij}| = \sqrt{3}), \quad \textbf{v}_i' \leftarrow \textbf{v}_i + f^V ( \textbf{v}_i, \sum_{j \in \mathcal{N}(i)} \textbf{e}_{ij}')\cdot \mathbb{I}(|\textbf{e}_{ij}'| = \sqrt{3}).
\label{coarse mp}
\end{equation}

Here $\textbf{e}_{ij}'$ and $\textbf{v}_i'$ denote the updated edge and node embeddings, respectively. $\textbf{v}_i$ represents the raw embedding of a target node, while $\textbf{v}_j$ represents the raw embedding of a neighboring node of $i$. $\mathbb{I}(|\textbf{e}_{ij}'| = \sqrt{3})$ filters for long edges in coarse processors. $f^W$ and $f^V$ are MLP layers.

The multiscale strategy not only increases the receptive field of each node but also enhances computational efficiency by reducing the number of edges in the coarse graph to less than 30\% of those in the original graph, thereby accelerating the training process. Unlike the MultiScale MeshGraphNets \cite{fortunato2022multiscale} approach, our proposed long-short-edge message passing does not require pre-computed fine and coarse graphs and two additional networks for exchanging information between these graphs, which allows easier implementation.

 A comparative analysis of loss functions is provided in Appendix~\ref{loss function comparison}, in which we found that the soft Dice BCE loss (originally defined in~\cite{galdran2022optimal} and described in Eq. (\ref{soft bce dice})), with $\mathcal{L}_{\text{BCE}}$ specified in Eq. (\ref{bce}) and $\mathcal{L}_{\text{Dice}}$ in (\ref{dice}), demonstrates the best overall performance on our dataset.

 \begin{equation}
     \mathcal{L_{\text{Soft BCE Dice}}} \left( \textbf{y}, \hat{\textbf{y}} \right) = \frac{N-n}{N} \mathcal{L_{\text{BCE}}} \left( \textbf{y}, \hat{\textbf{y}} \right) + \frac{n}{N}  \mathcal{L_{\text{Dice}}} \left( \textbf{y}, \hat{\textbf{y}} \right)
\label{soft bce dice}
\end{equation}

\section{Results}\label{results}

\begin{figure}[!t]
  \centering
  \includegraphics[width=\textwidth]{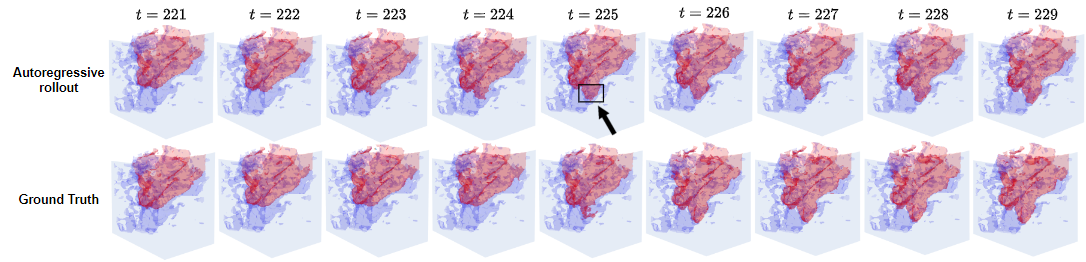}
  \caption{Visualisation (\href{https://www.youtube.com/watch?v=CyOpiv1anCY}{see video}) of ground truth and autoregressive rollout from $t = 221$ to 229 on the \textbf{testing set with emerging bubbles}. Red parts represent gas. Blue parts represent liquid.}
  \label{Validation 221-230}
\end{figure}

The training set comprises 20 time steps during which a bubble gradually emerges, alongside 20 additional steps with minor bubble oscillation. The model is evaluated to distinguish between these two scenarios on the testing set, which includes 8-time steps with emerging bubbles and 8-time steps with minor oscillation. The model is trained for 22.5 hours using one NVIDIA A100 GPU (80GB) graphics card. Stochastic gradient descent (SGD) with momentum was used for optimisation, with a learning rate of 0.001. Figure \ref{Validation 221-230} visualises the ground truth and the autoregressive rollout. The model not only accurately captures bubble emergence, but also predicts minor oscillations, as shown in Figures \ref{validation 288-295} and \ref{training 198-206}. 

\begin{equation}
\epsilon_{\tau} = \frac{|\hat{A}_{\tau} - A_{\tau}|}{A_{\tau}}
\label{surface_area_error}
\end{equation}

\begin{equation}
\bar{\epsilon} = \frac{1}{t_T-t_0} \sum_{\tau =t_0}^{t_T} \epsilon_{\tau}
\label{mean_surface_area_error}
\end{equation}

We define the mean surface area error, $\bar{\epsilon}$, as a metric to evaluate the performance of different loss functions, as shown in Eq. (\ref{mean_surface_area_error}). In this equation, $\tau$ denotes the time step involved in the prediction, ranging from the initial time step $t_0$ to the final time step $t_T$. The term $\epsilon_{\tau}$, defined in Eq. (\ref{surface_area_error}), represents the percentage surface area error at a given time step $\tau$. In this context, $\hat{A}_{\tau}$ denotes the surface area of prediction while $A_{\tau}$ denotes that of ground truth.The surface area calculation for each time step is detailed in Algorithm \ref{surface_area_calculation}. Table \ref{model_configuration_comparison} compares validation performance of different model configurations, indicating that F6C6F6 achieves the lowest mean surface area error with a reasonable training time.

%%\text{, where } \mathcal{L}_{\text{BCE}}(y, \hat{y}) &= - \sum \left(y_i \log \hat{(y_i)} + (1-y_i) \log \left(1 - \hat{y_i} \right)\right) \quad \mathcal{L}_{\text{Dice}}(y, \hat{y}) = 1 - \frac{2|y \cup \hat{y}|}{|y| + |\hat{y}|}

\begin{table}[htbp]
\small
  \caption{Comparison of validation performance on different model configurations based on training using a 20-time step dataset. The mean surface area error  (defined as Eq. (\ref{mean_surface_area_error}) was evaluated on a 9-time step test set. F$n$C$m$F$n$ denotes $n$ fine, $m$ coarse, followed by $n$ fine message-passing layers.}
  \label{model_configuration_comparison}
  \centering
  \begin{tabular}{ccccc}
    \toprule
      Metric / Configuration   & F9C0F9 & F7C4F7 & F6C6F6 & F4C10F4  \\
    \midrule
    Training time (seconds/epoch) & 24.88  & 21.72  & 20.02 & 16.74 \\
    Mean surface area error $\bar{\epsilon}$ (defined as Eq. (\ref{mean_surface_area_error})) & 10.40\% & 9.05\% & \textbf{8.81\%} & 10.10\%  \\
    
    \bottomrule
  \end{tabular}
\end{table}

\section{Conclusion}

We introduced a new paradigm for learning the physics of multiphase flow through porous media directly from experimental datasets. The LSE-MGN architecture successfully predicts the gradual emergence of bubbles and instances with minor oscillations, relying solely on local information. The results here sets the foundation for a promising direction in future pore-scale modeling.

% Future work could incorporate historical states along with transformer architectures to capture temporal information. Furthermore, the current model is trained on only a small section of the data. Section-based training could be employed to include more sections in the training set, enabling the locally learned model to be applied across the entire graph.

\bibliography{reference}

%%%%%%%%%%%%%%%%%%%%%%%%%%%%%%%%%%%%%%%%%%%%%%%%%%%%%%%%%%%%

\newpage
\appendix
% Optionally include supplemental material (complete proofs, additional experiments and plots) in appendix.
% All such materials \textbf{SHOULD be included in the main submission.}

\section{Pore structure}\label{pore_structure}
\begin{figure}[htbp]
  \centering
  \includegraphics[width=0.8\textwidth]{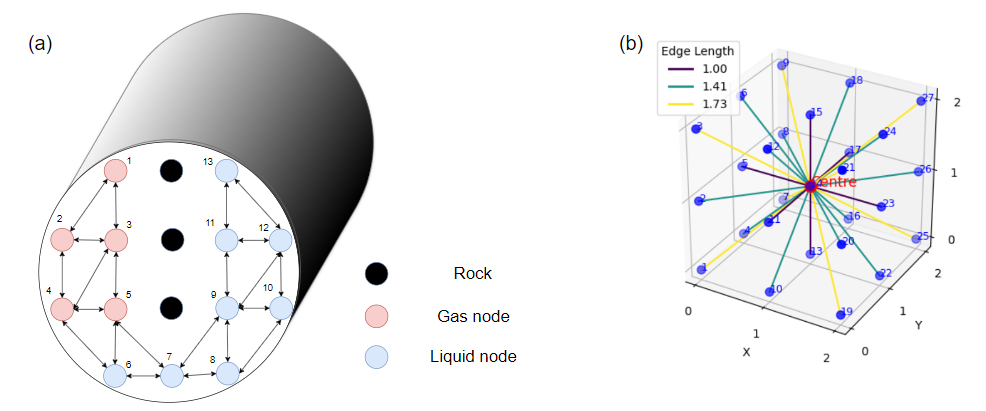}
  \caption{(\textbf{a}) An example of a cross section of a porous rock showing gas and liquid particles separated by solid rocks. (\textbf{b}) Edge formation: connect nodes within $\sqrt{3}$ units apart.}
  \label{connectivity}
\end{figure}

As illustrated in Figure \ref{connectivity} (a), node 1 cannot directly transmit information to node 13 due to the obstruction posed by the solid rock between them. Instead, the information must be relayed through a sequence of intermediate nodes, such as following the path $1 \rightarrow 3 \rightarrow 5 \rightarrow 7 \rightarrow 9 \rightarrow 11 \rightarrow 13$. This routing is necessary to bypass the blockage created by the rock. Figure \ref{connectivity} (b) illustrates how neighbours are defined during the graph construction process. Nodes are considered as neighbors if they are within a distance of $\sqrt{3}$ units from one another. Each node can connect to up to 27 neighbouring nodes. On average, within the selected section of the pore space, each node is connected to approximately 23.2071 neighbours.

% \section{Brief description of the data collection experiment}\label{experiment data}

% The experiments were conducted in a cylindrical carbonate sample, 5~mm diameter and 20~mm length. The sample was saturated with brine (deionised water doped with 15wt.\% KI to improve the X-ray contrast). Then nitrogen and brine were injected simultaneously. X-ray images were acquired to capture the transition from transient effects to steady-state flow. Each image took 2~s to acquire, and had a resolution of 2.75~$\mu m$. For a complete experimental description see \cite{spurin2020real}. Steady-state was determined by a stable pressure drop measured across the core. 

% The greyscale X-ray images were segmented to extract the location of the fluids present (gas and brine). This was done using the method described in \cite{spurin2023python}. 

\section{More visualisation of the prediction}\label{more_visulisation}

\begin{figure}[htbp]
  \centering
  \includegraphics[width=\textwidth]{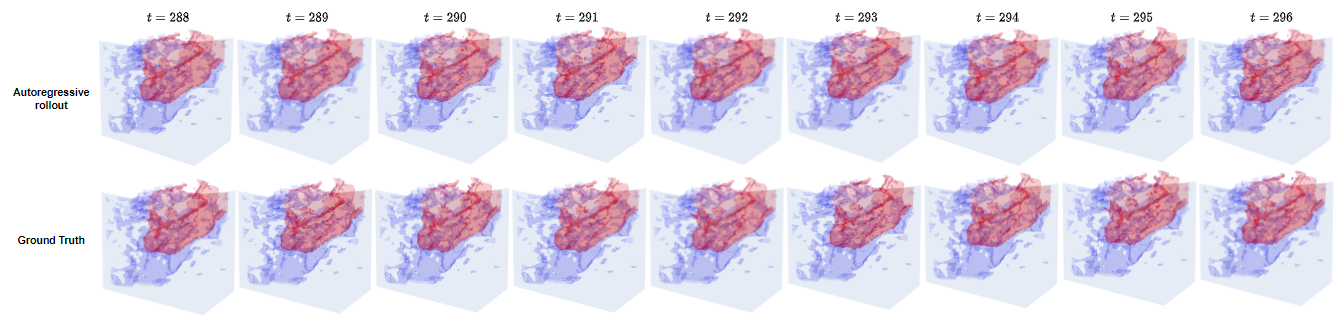}
  \caption{Visualisation (\href{https://www.youtube.com/watch?v=hbbWUJxL3ng}{see video}) of ground truth and autoregressive rollout from $t = 288$ to 296 on the \textbf{testing set with minor oscillations}. Red parts represent gas. Blue parts represent liquid.}
  \label{validation 288-295}
\end{figure}

\begin{figure}[htbp]
  \centering
  \includegraphics[width=\textwidth]{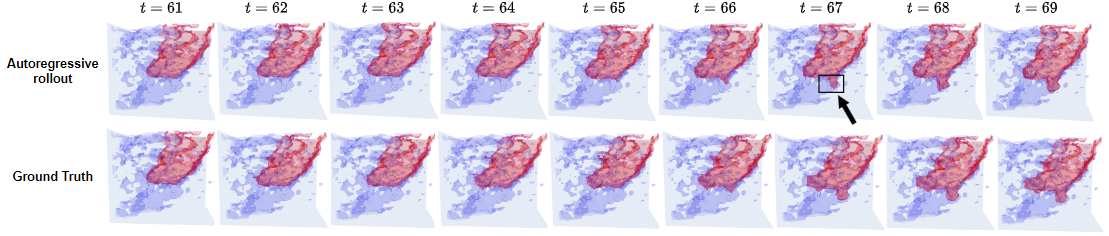}
  \caption{Visualisation (\href{https://www.youtube.com/watch?v=nm4u_G7pOcs}{see video}) of ground truth and autoregressive rollout from $t = 61$ to 69 on the \textbf{training set with emerging bubbles}. Red parts represent gas. Blue parts represent liquid.}
  \label{training 61-69}
\end{figure}

\begin{figure}[htbp]
  \centering
  \includegraphics[width=\textwidth]{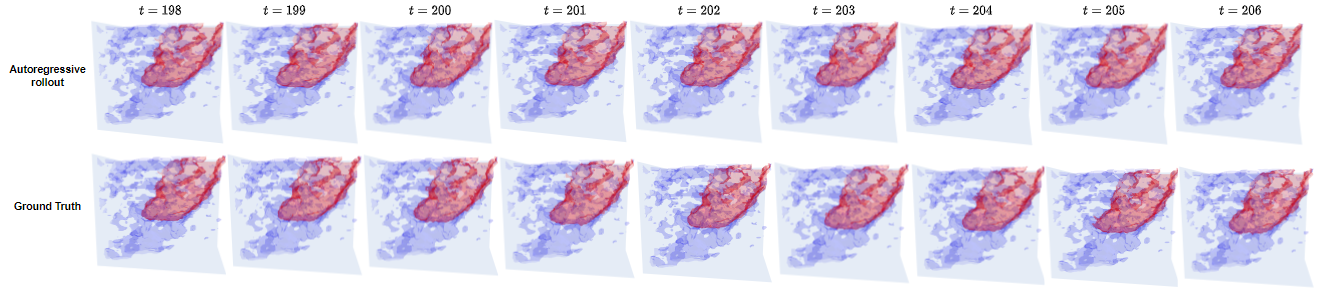}
  \caption{Visualisation (\href{https://www.youtube.com/watch?v=Rp_eI_iVqrY}{see video}) of ground truth and autoregressive rollout from $t = 198$ to 206 on the \textbf{training set with minor oscillations}. Red parts represent gas. Blue parts represent liquid.}
  \label{training 198-206}
\end{figure}

\section{Loss functions comparison}\label{loss function comparison}
We compared different loss functions mentioned in \cite{galdran2022optimal}\cite{caliva2019distance}\cite{lin2017focal}\cite{salehi2017tversky} on our training and validation sets . The training set includes 20 time steps with gradually emerging bubbles, while the validation set consists of 9 time steps with similar gradual bubble emergence. 

\begin{equation}
    \mathcal{L_{\text{BCE}}} \left( \textbf{y}, \hat{\textbf{y}} \right) = - \frac{1}{M} \sum_{i=1}^M \left[ y_i \text{log} \hat{y_i} + (1-y_i) \text{log} \left( 1 - \hat{y_i} \right) \right]
\label{bce}
\end{equation}

\begin{equation}
    \mathcal{L_{\text{Dice}}} \left( \textbf{y}, \hat{\textbf{y}} \right) = 1 - \frac{2|\textbf{y} \cup \hat{\textbf{y}}|}{|\textbf{y}| + |\hat{\textbf{y}}|} = 1 - \frac{2\langle \textbf{y}, \hat{\textbf{y}}\rangle}{\langle \textbf{y}, \textbf{y} \rangle + \langle \hat{\textbf{y}}, \hat{\textbf{y}}\rangle}
\label{dice}
\end{equation}

\begin{equation}
    \mathcal{L_{\text{Tversky}}} \left( \textbf{y}, \hat{\textbf{y}} \right) = 1 - \frac{2|\textbf{y} \cup \hat{\textbf{y}}|}{\alpha|\textbf{y}| + \beta|\hat{\textbf{y}}|} = 1 - \frac{\langle \textbf{y}, \textbf{y}\rangle}{\langle \textbf{y}, \textbf{y} \rangle  + \alpha \langle \textbf{y}, \textbf{1}-\hat{\textbf{y}} \rangle+ \beta\langle \textbf{1-y}, \hat{\textbf{y}}\rangle}
\label{Tversky}
\end{equation}

\begin{equation}
     \mathcal{L_{\text{Hard BCE Dice}}} \left( \textbf{y}, \hat{\textbf{y}} \right) = \begin{cases}
         \mathcal{L_{\text{BCE}}} \left( \textbf{y}, \hat{\textbf{y}} \right), \quad \text{if } n < 0.9N \\
          \mathcal{L_{\text{Dice}}} \left( \textbf{y}, \hat{\textbf{y}} \right), \quad \text{Otherwise}
     \end{cases}
\label{hard bce dice}
\end{equation}

\begin{equation}
     \mathcal{L_{\text{Distance map + BCE}}} \left( \textbf{y}, \hat{\textbf{y}} \right) = \frac{1}{N} \sum_{i=1}^N (1 + \Phi) \odot \mathcal{L_{\text{BCE}}} \left( \textbf{y}, \hat{\textbf{y}} \right)
\label{Distance map bce}
\end{equation}

\begin{equation}
     \mathcal{L_{\text{Distance map + Soft Dice BCE}}} \left( \textbf{y}, \hat{\textbf{y}} \right) = \frac{1}{N} \sum_{i=1}^N (1 + \Phi) \odot  \mathcal{L_{\text{Soft BCE Dice}}} \left( \textbf{y}, \hat{\textbf{y}} \right)
\label{Distance map soft dice bce}
\end{equation}

\begin{equation}
    \mathcal{L}_{\text{focal}} \left( \textbf{y}, \hat{\textbf{y}} \right) = -\sum_{i=1}^{N} \alpha_i \left(1 - \hat{y}_i \right)^{\gamma} y_i \log \hat{y}_i
\label{focal loss}
\end{equation}

Eq. (\ref{bce}) shows the Binary Cross-Entropy (BCE) loss, which measures the difference between two probability distributions but struggles with imbalanced data. Dice loss (see Eq. (\ref{dice})) addresses this imbalance by considering the intersection and union of ground truth and predictions, though it may be less accurate than BCE alone. Tversky loss (see Eq. (\ref{Tversky})) adjusts the weighting of false positives and false negative. Combining BCE and Dice loss, as in Eq. (\ref{soft bce dice}) and Eq. (\ref{hard bce dice}), improves performance by balancing accuracy and data imbalance.

The distance map indicates the distance of points with value 0 to the nearest point with value 1, serving as a weight to penalise points near the gas-liquid boundary more heavily. It can be combined with BCE and soft Dice BCE, as shown in Eq. (\ref{Distance map bce}) and (\ref{Distance map soft dice bce}). Additionally, focal loss (see Eq. (\ref{focal loss})) emphasises `hard' examples by reducing the loss contribution of easy samples.

\begin{table}[htbp]
  \caption{Comparison of mean surface area error $\bar{\epsilon}$ using different loss functions on training and testing sets. The surface area error for each time step is calculated as shown in Eq. (\ref{surface_area_error}). The reported error is the average across all time steps in autoregressive rollout (see Eq. (\ref{mean_surface_area_error})).}
  \label{loss_functions_surface_area}
  \centering
  \begin{tabular}{ccc}
    \toprule
    % \multicolumn{2}{c}{Part}                   \\
    % \cmidrule(r){1-2}
    Loss function     & $\bar{\epsilon}$ (Testing) & $\bar{\epsilon}$ (Training)  \\
    \midrule
    Soft dice bce &  \textbf{0.0988} & \textbf{0.0567} \\
    Hard dice bce &   0.1269 & 0.0941 \\
    Distance map + bce  &  0.1310 & 0.0840 \\
    Distance map + soft dice bce & 0.1550 & 0.1326\\
    Focal loss &  0.1122 & 0.0983 \\
    Tversky loss & 0.2086 & 0.2510\\
    Dice loss &  0.2120 & 0.2442\\
    BCE loss &  0.1502 & 0.1272 \\
    
    \bottomrule
  \end{tabular}
\end{table}

% \begin{equation}
%     \text{surface area error } \epsilon^{t_k} = \frac{|\text{Predicted surface area} - \text{Ground truth surface area}|}{\text{Ground truth surface area}}
% \label{surface_area_error}
% \end{equation}

 Table \ref{loss_functions_surface_area} compares the mean surface area error of various loss functions on training and validation sets. The training set consists of 20 time steps and the testing set consists of 9 time steps. The soft dice BCE loss achieves the lowest mean surface area error on both sets, making it the preferred choice in our work.

\begin{algorithm} 
\caption{Calculate Surface Area of a 3D Mesh in One Time Step} 
\label{surface_area_calculation} 
\begin{algorithmic}[1] 
\REQUIRE A 3D binary fluid dataset representing the spatial distribution of fluid within a volume. 
\ENSURE The total surface area of the generated 3D mesh.

\STATE Use the Marching Cubes algorithm to extract the 3D mesh from the input binary fluid data. This process computes a set of vertices $V = \{v_1, v_2, \ldots, v_n\}$ and a set of faces $F = \{f_1, f_2, \ldots, f_m\}$ that define the geometry of the mesh.

\STATE Initialise the total surface area: $A \gets 0$

\FOR{each face $f \in F$} 
\STATE Identify the three vertices $v_1, v_2, v_3$ that form the triangular face $f$. 

\STATE Compute the edge vectors: $\vec{a} \gets v_2 - v_1$ and $\vec{b} \gets v_3 - v_1$. 

\STATE Calculate the cross product of the edge vectors: $\vec{c} \gets \vec{a} \times \vec{b}$, which is perpendicular to the plane of the triangle. 

\STATE Compute the area of the triangle: $A_{\text{triangle}} \gets \frac{1}{2} \left\lVert \vec{c} \right\rVert$, where $\left\lVert \vec{c} \right\rVert$ is the magnitude of the cross product vector, giving twice the area of the triangle. 

\STATE Accumulate the triangle area into the total surface area: $A \gets A + A_{\text{triangle}}$ 

\ENDFOR
\RETURN $A$ \COMMENT{Return the total surface area of the 3D mesh}
\end{algorithmic} 
\end{algorithm}

\section{Links of visualisation videos}
Visualization videos for Figures 2 through 6 can be viewed by clicking the hyperlinks. If the YouTube links are not functional, the corresponding URLs are provided below for direct access.

\begin{itemize} 
    \item Figure \ref{Validation 221-230}: \url{https://www.youtube.com/watch?v=CyOpiv1anCY}
    \item Figure \ref{validation 288-295}: \url{https://www.youtube.com/watch?v=hbbWUJxL3ng}
    \item Figure \ref{training 61-69}: \url{https://www.youtube.com/watch?v=nm4u_G7pOcs}
    \item Figure \ref{training 198-206}: \url{https://www.youtube.com/watch?v=Rp_eI_iVqrY}
    \end{itemize}

\end{document}